\documentclass[floats,floatfix,showpacs,amssymb,prd,twocolumn,superscriptaddress,nofootinbib]{revtex4-1}

\usepackage{graphicx, epsf, epsfig, amssymb}
\usepackage{bm}
\usepackage{longtable}
\usepackage[usenames,dvipsnames]{xcolor}
\usepackage{color}
\usepackage[breaklinks]{hyperref}
\usepackage{amsfonts,amsmath,amssymb,mathrsfs}

\def\be{\begin{equation}}
\def\ee{\end{equation}}
\def\beq{\begin{eqnarray}}
\def\eeq{\end{eqnarray}}

\begin{document}

\title{Universality, maximum radiation and absorption\\in high-energy
       collisions of black holes with spin}
%
\author{Ulrich Sperhake}
\affiliation{Department of Applied Mathematics and Theoretical Physics,
Centre for Mathematical Sciences, University of Cambridge,
Wilberforce Road, Cambridge CB3 0WA, UK}
\affiliation{Institute of Space Sciences, CSIC-IEEC, 08193 Bellaterra,
Spain}
\affiliation{CENTRA, Departamento de F\'{\i}sica, Instituto Superior
  T\'ecnico, Universidade T\'ecnica de Lisboa - UTL, Avenida Rovisco Pais
  1, 1049 Lisboa, Portugal}
\affiliation{California Institute of Technology, Pasadena, CA 91125, USA}
\affiliation{Department of Physics and Astronomy, The University of
Mississippi, University, MS 38677, USA}

\author{Emanuele Berti}
\affiliation{Department of Physics and Astronomy, The University of
Mississippi, University, MS 38677, USA}
\affiliation{California Institute of Technology, Pasadena, CA 91125, USA}
\author{Vitor Cardoso}
\affiliation{CENTRA, Departamento de F\'{\i}sica, Instituto Superior
  T\'ecnico, Universidade T\'ecnica de Lisboa - UTL, Av.~Rovisco Pais
  1, 1049 Lisboa, Portugal}
\affiliation{Department of Physics and Astronomy, The University of
Mississippi, University, MS 38677, USA}
\author{Frans Pretorius}
\affiliation{Department of Physics, Princeton University, Princeton,
  NJ 08544, USA}


\date{\today}

\begin{abstract}
  We explore the impact of black hole spins on the dynamics of
  high-energy black hole collisions. We report results from numerical
  simulations with $\gamma$--factors up to $2.49$
  and dimensionless spin parameter
  $\chi=+0.85,\,+0.6,\,0,\,-0.6,\,-0.85$. We find
  that the scattering threshold becomes independent of spin at large
  center-of-mass energies, confirming previous conjectures that
  structure does not matter in ultrarelativistic collisions.  It has
  further been argued that in this limit all of the kinetic energy of
  the system may be radiated by fine tuning the impact parameter to
  threshold. On the contrary, we find that only about $60\%$ of the
  kinetic energy is radiated for $\gamma=2.49$. By monitoring apparent
  horizons before and after scattering events we show that the
  ``missing energy'' is absorbed by the individual black holes in the
  encounter, and moreover the individual black-hole spins change
  significantly.  We support this conclusion with perturbative
  calculations. An extrapolation of our results to the limit
  $\gamma\to \infty$ suggests that about half of the center-of-mass
  energy of the system can be emitted in gravitational radiation,
  while the rest must be converted into rest-mass and spin energy.
\end{abstract}

\maketitle

\noindent{\bf{\em I. Introduction.}}
Numerical relativity simulations have begun to shed light on
problems of fundamental interest in high-energy physics, such as
trans-Planckian scattering and gauge-gravity dualities
\cite{Cardoso:2012qm}. A scenario of particular interest in this
context is the collision of two black holes (BHs) near the speed of light,
which we will use here to test the validity of two key assumptions made
in Monte Carlo event generators
\cite{Dimopoulos2001,Cavaglia:2006uk,Dai:2007ki,Frost2009}
for the modeling of microscopic
BH production in trans-Planckian scattering~\cite{Arkani-Hamed1998,Randall1999}:
(i) that the spins of the colliding objects have a negligible effect
on the dynamics, and
(ii) that the mass of the formed BH is (up to a factor $\lesssim 1$)
given by the center of mass energy of the colliding particles,
i.e. that a significant fraction of the kinetic energy of the
system cannot be lost in the form of gravitational waves.
For this purpose
we perform a systematic analysis of $\sim 160$ collisions
of spinning and nonspinning BH binaries in $D=4$ (see
e.g.~\cite{Witek2010c,Okawa:2011fv} for early results in $D>4$)
to answer the following two questions: i) is the internal
structure of the colliding objects, here consisting of their
spin angular momentum, relevant? ii) is it possible (as suggested
in~\cite{Pretorius2007}) to radiate all of the kinetic energy in
fine-tuned encounters?

Our simulations answer both questions in the negative.
Spin effects become negligible for large $\gamma$: both the scattering
threshold and the maximum energy radiated become universal functions
of $\gamma$ (independent of spins).  For our largest boost
($\gamma=2.49$), grazing encounters radiate $\lesssim
60\%$ of the available kinetic energy. In fact this percentage
{\em decreases} with increasing boost velocity,
and barely varies with spin for $v\gtrsim 0.8$.
We show that the ``missing'' kinetic energy is
accounted for by an increase in the BH mass during the encounter.
These observations
justify the use of semianalytical calculations in classical general relativity
that neglect spins to
understand properties of the collisions,
and constrain the amount of
GWs radiated, which
will determine the initial mass spectrum of formed BHs.

Our results reinforce earlier evidence that ``matter does not
matter'': e.g. in \cite{Choptuik2009} it was shown
that collisions of two bosonic solitons at sufficiently high energies
lead to BH formation, and similar conclusions were reached when colliding
self-gravitating fluid objects \cite{Rezzolla:2012nr,East:2012mb}.
Compact fluids formed in head-on collisions of neutron stars
were also found to exhibit type I critical collapse in \cite{Jin:2006gm}.
\begin{figure*}[tb]
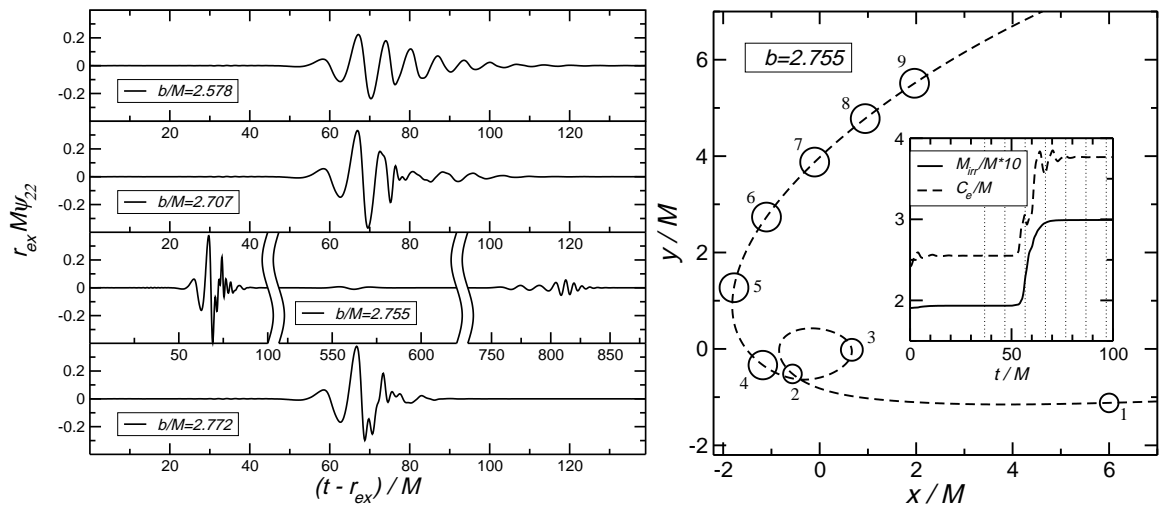

\begin{center}
\begin{tabular}{cc}
\epsfig{file=figure1a.eps,width=8.5cm,clip=true} 
\epsfig{file=figure1b.eps,height=6.6cm,clip=true} 
\end{tabular}
\caption{Left: Waveforms for $\gamma=2.49$, antialigned spins
  $\chi=0.6$ and
  selected values of $b$. The $b/M=2.755$ case is a triple encounter
  (two periastron passages followed by a merger). Right: Trajectory of
  one BH from the simulation with $b/M=2.755$. Inset: time
  evolution of the irreducible mass $M_{\rm irr}$ and of the
  circumferential radius $C_{\rm e}$ of each hole. The circles
  represent the BH location at intervals $\Delta t=10~M$
  (corresponding to vertical lines in the inset) and have radius equal
  to $M_{\rm irr}$.}
\label{fig:triple}
\end{center}
\end{figure*}

High-energy collisions of BHs
have been investigated extensively in $D=4$ spacetime
dimensions for equal-mass, nonspinning BHs, where the problem
is characterized by the boost factor
$\gamma=(1-v^2)^{-1/2}$ and impact parameter $b=L/P$, with $v$ 
the center-of-mass velocity, $L$ the initial orbital angular momentum
and $P$ the initial linear momentum of a single BH
(here and below we use geometrical units $G=c=1$).
In the head-on
case ($b=0$), high-energy BH collisions can radiate up to $14 \pm
3\,\%$ of the center-of-mass (CM) energy, and they
always produce a nonspinning remnant \cite{Sperhake:2008ga}. Grazing
collisions with $b\neq 0$, on the other hand, result in one of the
following three outcomes \cite{Sperhake:2009jz}: (i) a prompt merger
for small $b<b^*$, (ii) a ``delayed'' merger for $b^* \le
b < b_{\rm scat}$, or (iii) scattering of the holes to infinity for
$b\ge b_{\rm scat}$.  Here $b_{\rm scat}$ denotes the scattering
threshold and $b^* < b_{\rm scat}$ the ``threshold of immediate
merger'': by fine-tuning around $b^*$ a binary approaches a
near-circular orbit for a time $T\propto \log |b-b^*|$, before separating
or merging to form a single Kerr BH \cite{Pretorius2007}.  In $D=4$,
grazing collisions with $\gamma\leq 2.9$ were found to radiate as
much as $35\pm 5\,\%$ of the CM energy in gravitational waves (GWs)
and form BHs with near-, yet sub-extremal spins \cite{Sperhake:2009jz}.  A parallel study
investigated the scattering threshold $b_{\rm scat}$
for nonspinning binaries as a function of the CM energy
$M=\gamma M_0$ up to $\gamma=2.3$, finding $b_{\rm scat}\sim 2.5\,(M/v)$ \cite{Shibata2008}.
Comparisons with BH perturbation theory and point-particle
collisions in the zero-frequency limit provide a satisfactory
understanding of the qualitative features of these simulations
\cite{Pretorius2007,Berti2010,Gundlach:2012aj}, but several
outstanding questions remain.

\noindent{\bf{\em II. Setup.}}
%
Our simulations have been performed with the
{\sc Lean} code described in \cite{Sperhake:2006cy};
see also \cite{Cactusweb,
Schnetter:2003rb,Thornburg:1995cp,Thornburg:2003sf,
Ansorg:2004ds}.
We obtain stable evolutions by applying
two modifications to the numerical infrastructure
employed in our
previous studies
\cite{Sperhake:2008ga,Sperhake:2009jz, Sperhake:2010uv}: (i)
we evolve the conformal factor
as described in Sec.~II of \cite{Marronetti:2007wz}, and (ii) we
reduce the Courant factor to $0.45$.
The holes start on the $x$
axis with radial momentum $P_x$ and tangential momentum $P_y$,
separated by a distance $d$. The impact parameter is $b\equiv L/P =
P_y d/P$.
We extract gravitational radiation by computing the Newman-Penrose
scalar $\Psi_4$ at different radii $r_{\rm ex}$ from the center of the
collision.  $\Psi_4$ is decomposed into multipoles $\psi_{lm}$
as described in Ref.~\cite{Sperhake:2009jz}, but measuring the
polar angle $\theta$ relative to the $x$ axis.

Spurious ``junk'' radiation in the initial data is quite
insensitive to $b$, and comparable to our recent findings
\cite{Sperhake:2008ga,Sperhake:2009jz}. We remove it from reported
results in the same manner.  Errors due to discretization and
finite extraction radius are comparable to those reported in
\cite{Sperhake:2008ga,Sperhake:2009jz}. We estimate uncertainties in
radiated quantities of $3~\%$ and $15~\%$ for low and high boosts,
respectively. These are dominated by discretization errors in the
wave zone, which may be addressed in future work using
multi-patch techniques \cite{Sarbach:2012pr}.

Contrary to our recent investigation of ultrarelativistic encounters of
spinning BHs in ``superkick'' configurations \cite{Sperhake:2010uv}, here
we expect the dynamics to be most strongly affected by
the ``hang-up'' effect~\cite{Campanelli:2006uy} typical of spins
(anti)aligned with the orbital angular momentum {\boldmath$L$}.
We study its boost dependence by evolving three sequences of
equal-mass BH binaries: (i) a sequence with zero spins, (ii) a sequence
with dimensionless spin parameters $\chi=\chi_1=\chi_2=0.6$ aligned with
{\boldmath$L$}, and (iii) a sequence with spins $\chi=0.6$
antialigned with
{\boldmath${L}$}. For each sequence we
consider four boost parameters ($\gamma=1.22,\,1.42,\,1.88,\,2.49$)
and for each $\gamma$ we simulate encounters with about 10 different
values of $b$ to bracket the scattering threshold.
In addition, we study the boost values $\gamma=1.22,\,2.49$
in the same manner using larger spins $\chi=0.85$ aligned or
anti-aligned with {$\boldmath{L}$}.

\noindent{\bf{\em III. Scattering threshold.}}
We expect a given initial binary configuration to result in either
a prompt merger, a delayed merger or scattering to infinity.  Our new
simulations confirm this scenario. This is illustrated in the left panel
of Fig.~\ref{fig:triple}, where we plot a subset of representative
waveforms from the $\gamma=2.49$ sequence with antialigned spins $\chi=0.6$.
For small impact parameter (top) the BHs merge promptly, and the signal
is a clean merger/ringdown waveform, leading to formation of a BH with
dimensionless spin $\chi_{f}\simeq 0.87$.  For the second waveform from
the top the merger is not quite prompt: it shows a pattern similar to
the scattering waveforms with $b/M=2.772$ (bottom panel), followed by
a ringdown.
The third waveform for $b/M=2.755$ is a rare {\em triple} encounter
consisting of two revolutions (the second close encounter is visible
as a small ``bump'' at $t/M\sim 550$), followed by a merger signal with
relatively low amplitude.  Note that the binary radiates and partially
absorbs much of the system's kinetic energy during the first
encounter, therefore subsequent encounters occur at low velocity and
radiate much less. We display this behaviour in the right
panel of Fig.~\ref{fig:triple}, where we plot the trajectory of one BH
for the configuration $b/M=2.755$ and represent snapshots (labeled `1' to
`9') of the BH at time intervals $\Delta t=10~M$ by circles with radius
equal to the irreducible mass $M_{\rm irr}$. From snapshots `2' to `4' we observe a
rapid increase in the ``size'' of the black hole; successive snapshots
are located closer to each other, showing that the BH has slowed down
as a result of the interaction.

In order to determine $b_{\rm scat}$ as a function of spin and boost
we need to distinguish between merging and scattering
collisions. Mergers are easily identified by finding a common AH.  We
identify an encounter as a scattering case when the following criteria
are met: (i) no common AH is found; (ii) the Kretschmann scalar
(defined in terms of the Riemann tensor as
$R_{\alpha \beta \gamma \delta} R^{\alpha \beta \gamma \delta}$) at the
origin approaches zero at late times within numerical uncertainties;
and (iii) the coordinate trajectories of the BHs, Eq.~(14) in
Ref.~\cite{Sperhake:2006cy}, separate out to values comparable to
their initial distance.
\begin{figure}[t]
\begin{center}
\begin{tabular}{cc}
\epsfig{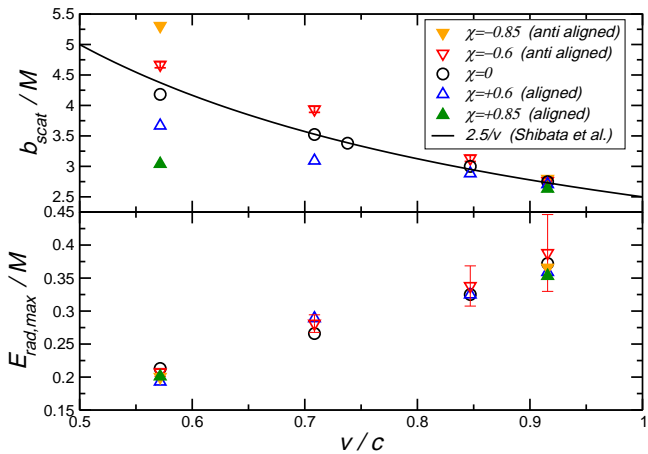} &
\end{tabular}
\caption{Critical scattering threshold (upper panel) and maximum
  radiated energy (lower panel) as a function of $v$.
  Colored ``triangle'' symbols pointing up and down
  refer to the aligned and antialigned cases,
  respectively. Black ``circle'' symbols represent the thresholds for
  the four nonspinning configurations studied in this paper,
  complemented (in the upper panel) by results from
  \cite{Sperhake:2009jz} for $v=0.753$. For clarity, we only plot
  error bars for the antialigned-spin sequence; for
  $b_{\rm scat}$ they
  are comparable in size to the symbols.
  \label{fig:d4bscat}}
\end{center}
\end{figure}

The scattering thresholds obtained in this way are plotted in the
upper panel of Fig.~\ref{fig:d4bscat}.  Errors in $b_{\rm scat}$
come from numerical truncation error and discrete sampling of the
parameter space, estimated as follows. In the most challenging case
($\gamma=2.49$) our standard-resolution runs with grid spacing $\Delta
x$ yield $b_{\rm scat}/M=2.760$ for the antialigned case with
$\chi=0.6$. By running
simulations at two higher resolutions $0.9~\Delta x$ and $0.8~\Delta x$,
we find $b_{\rm scat}/M=2.741$ and $2.731$ respectively, corresponding to
about fourth-order convergence, a Richardson-extrapolated value $b_{\rm
scat}/M=2.713$, and therefore a numerical uncertainty of $0.047$. The
error due to discretization of the parameter space ($\sim 1.6\times
10^{-3}$) is negligible in comparison, so we adopt $\delta b_{\rm
scat}/M \approx 0.05$ as a conservative error estimate.
Fig.~\ref{fig:d4bscat} shows that the scattering threshold is
spin-independent in the limit $\gamma \to \infty$.  Our nonspinning
simulations are consistent with the results obtained by Shibata et
al. \cite{Shibata2008} at lower boosts and within $\sim 50~\%$ of the
shock-wave analysis in Table II of \cite{Yoshino:2005hi}, that suggests
$b_{\rm scat}/M\gtrsim 1.68$ in the ultrarelativistic limit.

\noindent{\bf{\em IV. Maximum radiation.}}
As pointed out in \cite{Sperhake:2009jz,Gold:2012tk}, the total energy
$E_{\rm rad}/M$ radiated in grazing BH collisions increases steeply as
the impact parameter approaches $b^*$ or $b_{\rm scat}$. This is true
also for spinning binaries.  For $\gamma = (1.22,\,1.42,\,1.88,\,2.49)$,
respectively, we find the maxima in $E_{\rm rad}/M$ plotted in the lower
panel of Fig.~\ref{fig:d4bscat}.
For reference, the initial fraction of total energy in the form of
kinetic energy $K/M=(\gamma-1)/\gamma$ is $(17.9,29.4,46.8,59.8)\%$,
and for $\chi=0.6$ we have $(4.2,3.6,2.6,1.7)\%$ in initial
spin energy.

\begin{table}[t]
  \begin{ruledtabular}
  \begin{tabular}{cccccccc}
Spin & $\gamma$ &$b/M$ &$K/M$ &$E_{\rm rad}/K$ 
&$E_{\rm abs}/K$ &$|\chi_{\rm i}|$ &$|\chi_{\rm s}|$\\
\hline
$\downarrow 0.85$ & 1.22 &5.322 & 0.179 &0.870 &0.088 &0.84 &0.66 \\
$\downarrow 0.85$ & 2.49 &2.784 & 0.598 &0.602 &0.329 &0.82 &0.12 \\
\hline
$\downarrow 0.6$  & 1.22 &4.671 & 0.179 &0.899 &0.088 &0.60 &0.45 \\
$\downarrow 0.6$  & 1.88 &3.133 & 0.468 &0.665 &0.273 &0.59 &0.01 \\
$\downarrow 0.6$  & 2.49 &2.762 & 0.598 &0.570 &0.313 &0.57 &0.12 \\
\hline
0 & 1.22 &4.191 & 0.179 &0.899 &0.075 &0.01 &0.06 \\
0 & 1.88 &3.005 & 0.468 &0.637 &0.284 &0.08 &0.24 \\
0 & 2.49 &2.749 & 0.598 &0.574 &0.320 &0.10 &0.23 \\
\hline
$\uparrow 0.6$  & 1.22 &3.678 & 0.179 &0.894 &0.065 &0.60 &0.59 \\
$\uparrow 0.6$  & 1.88 &2.886 & 0.468 &0.618 &0.284 &0.59 &0.45 \\
$\uparrow 0.6$  & 2.49 &2.704 & 0.598 &0.600 &0.320 &0.57 &0.33 \\
\hline
$\uparrow 0.85$ & 1.22 &3.053 & 0.179 &0.875 &0.053 &0.84 &0.80 \\
$\uparrow 0.85$ & 2.49 &2.643 & 0.598 &0.557 &0.292 &0.82 &0.39 \\
  \end{tabular}
  \end{ruledtabular}
  \caption{\label{UliAbsorption} Fractional kinetic energy radiated
    and absorbed for representative simulations with aligned
    spins ($\uparrow$), antialigned spins ($\downarrow$) or
    nonrotating BHs (`0'). We also list spin estimates $\chi_{\rm i}$
    and $\chi_{\rm s}$ before and after
    the encounter. $\chi_i$ is
    measured at a time $\sim 20~M$ after the beginning of the
    simulation. Small deviations from the initial data parameter
    $\chi=\pm0.85,\pm0.6,0$
    can presumably be attributed to the BHs absorbing
    an increasing amount of junk radiation as $\gamma$ increases.}
\end{table}

For a subset of scattering runs where we monitored the apparent
horizon as a function of time, in Table~\ref{UliAbsorption} we list
estimates of radiated energies and spins before/after the first
encounter.  These numbers reveal two striking features: (i) the
maximum radiated energy varies mildly with spin at any given $\gamma$,
and (ii) for small boosts the maximum radiation is comparable to the
initial kinetic energy; however as $\gamma$ increases the ratio drops,
down to $\sim 60\%$ for $\gamma=2.49$.  This observation prompts two
questions. Where has the remaining kinetic energy gone?  Why does the
deficit increase with boost?

\noindent{\bf{\em V. Absorption.}}
The answer to these questions is found in the apparent horizon dynamics
of the individual holes before and after the first encounter. We have
analyzed the data in detail for a set of
binary configurations where the
individual holes separate sufficiently after first encounter to warrant
application of the isolated horizon limit.  Specifically, we measure
the equatorial circumference $C_{\rm e}=4\pi M$ and the irreducible mass
$M_{\rm irr}$ of each BH before and after the encounter. The inset of
the right panel of Fig.~\ref{fig:triple} shows the variation of these
quantities with time in a typical simulation: absorption occurs
over a short timescale $\approx 10M$.
Since the apparent horizon area $A_{\rm AH}=16\pi M_{\rm
  irr}^2=[C_{\rm e}^2/(2\pi)](1+\sqrt{1-\chi^2})$, in this way we can
estimate the rest mass and spin of each hole before ($M_{\rm i}$,
$\chi_{\rm i}$) and after ($M_{\rm s}$, $\chi_{\rm s}$) the first
encounter. We define the absorbed energy $E_{\rm abs}=2(M_s-M_i)$.
The results in Table \ref{UliAbsorption} show that the sum $(E_{\rm
rad}+E_{\rm abs})/M$ accounts for most of the total available kinetic
energy in the system, and therefore the system is no longer kinetic-energy
dominated after the encounter. A fit of the data yields $E_{\rm
rad}/K=0.46 (1 + 1.4/\gamma^2)$ and $E_{\rm abs}/K=0.55(1-1/\gamma)$,
suggesting that radiation and absorption contribute about equally in
the ultrarelativistic limit, and therefore that absorption sets an upper
bound on the maximum energy that can be radiated.

The fact that absorption and emission are comparable in the
ultrarelativistic limit is supported by point-particle calculations in
BH perturbation theory. For example, Misner et
al. \cite{Misner:1972jf} studied the radiation from ultrarelativistic
particles in circular orbits near the Schwarzschild light ring,
i.e. at $r=3M(1+\epsilon)$. Using a scalar-field model they found that
50\% of the radiation is absorbed and 50\% is radiated as $\epsilon\to
0$. We verified by an explicit calculation ignoring self-force effects
that the same conclusion applies to {\em gravitational} perturbations
of Schwarzschild BHs (cf.~\cite{Breuer:1973kt}). A recent analysis
including self-force effects finds that $42\%$ of the energy should be
absorbed by nonrotating BHs as $\epsilon\to 0$ (cf.~Fig.~4 in
\cite{Gundlach:2012aj}).

Rather than considering particles near the light ring, we can model
our problem using particles plunging ultrarelativistically into (for
simplicity) a Schwarzschild BH. Davis et al. \cite{Davis:1972ud} first
computed the energy absorbed when a particle of mass $m$ falls {\em
  from rest} into a Schwarzschild BH of mass $M_{\rm BH}$. Remarkably,
they found that the total absorbed energy (summed over all multipoles
$\ell\geq 2$) diverges. This is due to the fact that most of the
absorption occurs near the horizon, so we must go beyond the
point-particle approximation and introduce a physical cutoff at
$\ell_{\rm max}\approx \pi M_{\rm BH}/2m$ to take into account the
finite size of the infalling particle. For comparable-mass encounters
it is reasonable to truncate the sum at $\ell=2$.
By adapting the BH perturbation theory code of \cite{Berti:2010gx}, we
extended the calculation of \cite{Davis:1972ud} to
generic particle energies $p_0=E/m$.
Our calculation shows that the radiated (absorbed) energy is $E_{\rm
  rad,abs}^{\rm PP}=k_{\rm rad,abs} (p_0 m)^2/M_{\rm BH}$, with
$k_{\rm rad}=(1.04\times 10^{-2},3.52\times 10^{-2},0.119,0.262)$
and $k_{\rm abs}=(0.304,0.310,0.384,0.445)$ for $p_0=(1,1.5,3,100)$,
respectively. So, again, in the ultra-relativistic
limit the point-particle model
predicts a roughly comparable amount of emission and absorption. As a
consequence of this significant energy absorption, in the large-$\gamma$
limit close scattering encounters between two arbitrarily small (in rest
mass) BHs can result in two, slowly moving BHs with rest mass increased
by a factor of order $\gamma$.

Another remarkable implication of Table \ref{UliAbsorption} is that
high-energy scattering encounters can significantly modify the
spin magnitude.  For example, when $\gamma=2.5$ the absolute value
of the BH spin decreases
from $\sim 0.6$ to $\sim 0.3$ in the aligned case, from $\sim 0.6$ to
$\sim 0.1$ in the antialigned case, and we measure a post-scattering
spin $\chi \sim 0.2$ for initially nonspinning encounters. These changes
in dimensionless spin parameter correspond to roughly the {\em same}
total angular momentum being transferred to the BHs during the
interaction, independent of the initial spin.

\noindent{\bf{\em VI. Ultrarelativistic extrapolation.}}
We estimate the maximum radiated energy when $\gamma \to \infty$ as
follows.  According to Ref.~\cite{Sperhake:2008ga}, head-on collisions
($b=0$) radiate at most $E_{\rm rad}/M\sim 0.14$ in this limit.
For each $\gamma$, the increase in radiation induced by fine-tuning
near threshold can be characterized by the ratio ${\cal R}\equiv
E_{\rm rad}(b=0)/E_{\rm rad}^{\rm max}$.  For our largest $\gamma$
we find ${\cal R}\sim 0.2$, and a fit to our data yields ${\cal
R}(\gamma)=0.34(1-1/\gamma)$.  Using this fit in combination with results
from Ref.~\cite{Sperhake:2008ga} we find $E_{\rm rad}^{\rm max}/M\approx
0.14/0.34\sim 0.41$ as $\gamma\to \infty$.  A more conservative upper
bound on $E_{\rm rad}^{\rm max}$ can be obtained noting that ${\cal R}$
increases with $\gamma$, but using the last data point in our simulations
as a lower limit on ${\cal R}$: this yields a limit $E_{\rm rad}^{\rm
max}/M\lesssim 0.14/0.2=0.7$ as $\gamma\to \infty$.
These results are consistent (within the errors) with our discussion of
Table~\ref{UliAbsorption}, which indicates that radiation in high-energy
encounters accounts for roughly 0.46 of the available energy. Our
simulations thus settle the long-standing question of whether it is
possible to release all of the CM energy as GWs in high-energy BH
collisions: the answer is no.

Two crucial assumptions underlie the study of BH production from
trans-Planckian particle collisions.  The first assumption, that BHs
are indeed produced in the collision, is on a firmer footing due to the
results of~\cite{Choptuik2009,Rezzolla:2012nr,East:2012mb}, where the hoop
conjecture was found to be valid even in highly dynamical situations.
The present study addresses the second crucial assumption, i.e., that
the internal structure of the colliding bodies is irrelevant at high
energies. Furthermore our simulations provide strong evidence that,
because of absorption, the maximum radiation produced in ultrarelativistic
encounters in four dimensions cannot exceed $\approx 50\%$ of the
CM energy.

\noindent{\bf \em Acknowledgements.}
E.B.'s research is supported by NSF CAREER Grant No. PHY-1055103.
F.P acknowledges support from the NSF through CAREER Grant PHY-0745779,
FRG Grant 1065710, and the Simons Foundation.
U.S. acknowledges support from the Ram\'on y Cajal Programme and Grant
FIS2011-30145-C03-03 of the Ministry of Education and Science of
Spain, the NSF TeraGrid and XSEDE Grant No. PHY-090003, RES Grant
Nos. AECT-2012- 1-0008 and AECT-2012-3-0012 through the Barcelona
Supercomputing Center and CESGA Grant Nos. ICTS- 200 and ICTS-221.
This work was supported by the DyBHo--256667 ERC Starting Grant, the
CBHEO293412 FP7-PEOPLE-2011-CIG Grant, the NRHEP 295189
FP7-PEOPLE-2011-IRSES Grant, and by FCT-Portugal through projects
PTDC/FIS/098025/2008, PTDC/FIS/098032/2008, PTDC/FIS/116625/2010 and
CERN/FP/116341/2010, by the Sherman Fairchild Foundation to
Caltech, and STFC and BIS through the DiRAC HPC initiative for the
Cosmos system at the University of Cambridge.

\bibliographystyle{h-physrev4}

\end{document}